# A Novel Wide-Area Control Strategy for Damping of Critical Frequency Oscillations via Modulation of Active Power Injections

Ruichao Xie, Innocent Kamwa, *Fellow, IEEE*, and C. Y. Chung, *Fellow, IEEE*

*Abstract*—This paper proposes a novel wide-area control strategy for modulating the active power injections to damp the critical frequency oscillations in power systems, this includes the inter-area oscillations and the transient frequency swing. The proposed method pursues an efficient utilization of the limited power reserve of existing distributed energy resources (DERs) to mitigate these oscillations. This is accomplished by decoupling the damping control actions at different sites using the oscillation signals of the concerned mode as the power commands. A theoretical basis for this decoupled modulating control is provided. Technically, the desired sole modal oscillation signals are filtered out by linearly combining the system-wide frequencies, which is determined by the linear quadratic regulator based sparsity-promoting (LQRSP) technique. With the proposed strategy, the modulation of each active power injection can be effectively engineered considering the response limit and steady-state output capability of the supporting device. The method is validated based on a two-area test system and is further demonstrated based on the New England 39-bus test system.

## I. INTRODUCTION

Poorly-damped inter-area oscillation modes limit the power transfer of a power system and may cause large-scale blackout. Traditional approaches for mitigation of these low-frequency oscillations include reducing the power flow of key transmission paths thru generation re-dispatch [1]; installing power system stabilizer (PSS) on synchronous machines [2] or supplementary damping controller on FACTS [3]. The common low-frequency mode associated with primary frequency response may be lightly damped; consequently, under sudden imbalance in generation and load, large transient frequency and power swings and even sustained frequency and power oscillations may occur, this may trigger load rejection, critical line tripping, and subsequent cascading issues. Installing Multi-Band PSSs [2], retuning the PID-type speed governors [4] and disabling the speed governors [5] are effective measures in practice for mitigating such swings.

This work was supported in part by the Natural Sciences and Engineering Research Council (NSERC) of Canada, and Hydro-Québec. *(Corresponding author: Innocent Kamwa)*

R. Xie is with the Department of Electrical and Computer Engineering, Université Laval, Quebec, QC G1V 0A6, Canada (e-mail: ruichao.xie.1@ulaval.ca).
I. Kamwa is with Hydro-Québec/IREQ, Varennes, QC J3X 1S1, Canada (e-mail: kamwa.innocent@ireq.ca).
C. Y. Chung is with the Department of Electrical and Computer Engineering, University of Saskatchewan, Saskatoon, SK S7N 5A9, Canada (e-mails: c.y.chung@usask.ca).

Recently, it has been demonstrated that modulating the active power injection of a HVDC to damp the critical inter-area oscillation of a large-scale power system is technically feasible [6]. As a matter of fact, with the proliferation of distributed energy resources (DERs), e.g., energy storage systems, active power modulation is becoming a cheap means of controlling the oscillatory dynamics of a power system. Intuitively, geographically dispersed actuators increase the controllability for the system-wide frequency oscillations while reducing the control burden at each single control site. However, in the context of classical feedback control, this may also significantly increase the computational burden brought about by the coordination of numerous controllers.

To date, major efforts toward the so-called multi-point active power modulation based damping control are focused on coordinating the control actions using output feedback control strategies. By using the local frequency as the power command, a design method of structurally constrained output feedback with bounded power responses is proposed in [7]; a non-linear simulation based optimization approach for coordinating the power modulations to damp the critical mode of oscillation is proposed in [8]; likewise, a gain tuning approach of load modulation for primary frequency regulation while considering the load's disutility is presented in [9]. The other method employs the system-wide frequencies to drive the active power modulations, a structurally constrained output feedback optimal control method is proposed in [10] for suppressing the inter-area oscillation of a two-area system; the approach is then further developed in [11] for controlling multiple inter-area modes while optionally improving the primary frequency response of a large-scale power system.

All of the above-mentioned methods have distinct merits. Nevertheless, we note that the control performance of the local frequency based power modulation may be dependent on the system structure and the actuator location. We also note that although all the methods pursue an optimal coordination, the strategies may not be very cost-effective in terms of utilizing the valued active power response to resolve the oscillation issue, which is usually dictated by a couple of modes under an operating condition; moreover, the approaches are based on a centralized implementation (i.e., the modulations act as a single control action), which may yield a compromised control effect, and it is not easy to illustrate the role of each single-point control action in the oscillation damping.

A design concept of DER-based primary frequency control is presented in [12], where the authors employ a reduced-order model to superimpose the frequency controllers to achieve a desired damping ratio of the common mode. Because the controllers are designed in a superposition manner, each power modulation can be conveniently engineered based on many considerations such as the response limit and steady-state output capability of the supporting device. Motivated by the result, this paper formulates a new output feedback control strategy so as to efficiently utilize the limited power reserve of existing active power injection assets to damp the critical inter-area oscillations and the transient frequency swing. The technical developments presented in this paper may be traced as follows:

1) A new control strategy for decoupling the damping control actions at different sites. The concept of modal decomposition control is revisited in the context of static feedback control, it is demonstrated that the distributed static feedback controllers that exclusively add damping to the same mode can be simply superimposed. This means that the control actions can be decoupled by using the sole modal oscillation signals as the input signals of the controllers. Moreover, by properly choosing control sites (areas), the decoupling feature may be retained for the control of multiple modes.

2) Flexible and efficient active power modulating control. With the proposed control strategy, the active power modulations can be flexibly engineered to efficiently utilize the limited power reserve of the supporting devices to mitigate the concerned mode of oscillation; the power response of each device tends to be smaller as more devices are engaged, thereby relieving the concern about the availability of the devices.

3) The linear quadratic regulator based sparsity-promoting (LQRSP) technique is formalized as a tool for determining the least number of system state variables that need to be combined for filtering out the desired sole modal oscillation signals. Extensive studies show that these states are the frequencies scattered in the system.

## II. Review of Modal Decomposition Control

In [13], the modal decomposition control is introduced for PSS design. In this section, the concept is reviewed in the context of static feedback and therefore provides a basis for the control strategy to be proposed in the next section.

A power system can be described by a set of differential and algebraic equations, from which a single-input single-output (SISO) model may be obtained by linearizing the system around an operating point as follows:

$$\dot{x} = Ax + B_i u \quad (1a)$$
$$y_i = C_i x \quad (1b)$$

where $x$ is an $n \times 1$ state vector, $A$ is an $n \times n$ state matrix, and $B_i$ and $C_i$ are the $n \times 1$ input and $1 \times n$ output vectors, respectively. The modal properties are given by $AM = M\Lambda$, $N^T M = I$, where $M$ and $N$ represent the right and left modal matrices, respectively, $I$ is an $n \times n$ identity matrix, and $T$ denotes transpose. $M = [m_1, m_2, \ldots, m_n]$, $N = [n_1, n_2, \ldots, n_n]$, where $m_j$ and $n_j$ are the $n \times 1$ right and left eigenvectors for mode $j$, respectively, and $\Lambda$ is a diagonal matrix comprised of the system eigenvalues, denoted by $\Lambda = \text{diag}(\lambda_1, \lambda_2, \ldots, \lambda_n)$.

Apply the linear transformation
$$x = Mz \quad (2)$$
where $z$ is the state vector in the new coordinates. Then,
$$\dot{z} = \Lambda z + N^T B_i u \quad (3a)$$
$$y_i = C_i M z = \sum_{j=1}^n C_i m_j z_j \quad (3b)$$

where $z_j$ is the $j^{th}$ component of $z$. Consider a sole modal signal scaled by a gain $K$ as control input, denoted as
$$u = K C_i m_j z_j \quad (4)$$

The state equations then become
$$\dot{z} = \Lambda z + N^T B_i K C_i m_j z_j = A^* z \quad (5)$$

where $A^*$ is the closed-loop system state matrix in the new coordinates and

$$A^* = \begin{bmatrix} \lambda_1 & 0 & \cdots & n_1^T B_i \; KC_i m_j & \cdots & 0 \\ 0 & \lambda_2 & \cdots & n_2^T B_i KC_i m_j & \cdots & 0 \\ \vdots & \vdots & \ddots & \vdots & \ddots & \vdots \\ 0 & 0 & \cdots & \lambda_j + n_j^T B_i KC_i m_j & \cdots & 0 \\ \vdots & \vdots & \cdots & \vdots & \ddots & 0 \\ 0 & 0 & \cdots & n_n^T B_i KC_i m_j & \cdots & \lambda_n \end{bmatrix} \quad (6)$$

The eigenvalues of the closed-loop system can be obtained by solving
$$\det(sI - A^*) = 0 \quad (7)$$

Equations (6) and (7) illustrate that the feedback control loop will only alter the $j^{th}$ mode. Ideally, if $n_j^T B_i K C_i m_j$ is a negative real number, the $j^{th}$ mode will be horizontally shifted to the left half plane without changing its frequency.

## III. The Proposed Control Strategy and its Application to Multi-Point Active Power Modulation

### A. The Proposed Control Strategy

Based on the theory in Section II, using the sole modal signal to close a static feedback control loop may be modeled as

$$u_d = P_d \sum_q K_{d,q} C_q m_j z_j$$
$$O_d = P_d \sum_q K_{d,q} C_q m_j \quad (8)$$

where $K_{d,q}$ denotes a gain for a modal signal of mode $j$ in system output $q$, and $P_d$ denotes a gain for the linearly combined sole modal signals. The input vector for the system input is $B_d$. Here, a system input (#d) denotes a controller. The controller is depicted in Fig. 1.

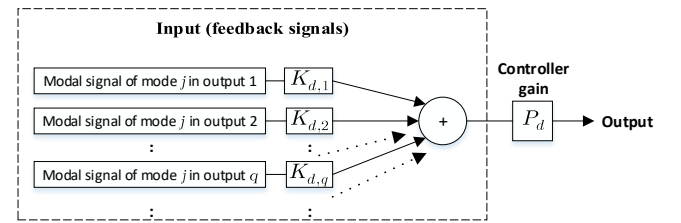

Fig. 1. The controller for the $j^{th}$ mode at system input (#d).

For the sake of illustration, consider that the $j^{th}$ mode is controlled by two controllers, i.e., the controllers at $u_1$ and $u_2$ access the modal signal of the $j^{th}$ mode, the closed-loop system state matrix in the new coordinates is

$$\boldsymbol{A}^* = \begin{bmatrix} \lambda_1 & 0 & \cdots & \boldsymbol{n}_1^T\boldsymbol{B}_1\boldsymbol{O}_1 + \boldsymbol{n}_1^T\boldsymbol{B}_2\boldsymbol{O}_2 & \cdots & 0 \\ 0 & \lambda_2 & \cdots & \boldsymbol{n}_2^T\boldsymbol{B}_1\boldsymbol{O}_1 + \boldsymbol{n}_2^T\boldsymbol{B}_2\boldsymbol{O}_2 & \cdots & 0 \\ \vdots & \vdots & \ddots & \vdots & \ddots & \vdots \\ 0 & 0 & \cdots & \lambda_j + \boldsymbol{n}_j^T\boldsymbol{B}_1\boldsymbol{O}_1 + \boldsymbol{n}_j^T\boldsymbol{B}_2\boldsymbol{O}_2 & \cdots & 0 \\ \vdots & \vdots & \cdots & \vdots & \ddots & 0 \\ 0 & 0 & \cdots & \boldsymbol{n}_n^T\boldsymbol{B}_1\boldsymbol{O}_1 + \boldsymbol{n}_n^T\boldsymbol{B}_2\boldsymbol{O}_2 & \cdots & \lambda_n \end{bmatrix} \quad (9)$$

It can be seen from (9) that the controllers will only alter the $j^{th}$ mode and the modified mode is given by $\lambda_j + \boldsymbol{n}_j^T\boldsymbol{B}_1\boldsymbol{O}_1 + \boldsymbol{n}_j^T\boldsymbol{B}_2\boldsymbol{O}_2$, which implies that the damping control actions are fully decoupled and therefore: each controller is allowed to independently set up its contribution to the mode by adjusting the gain $P$; adding or missing any controller will not affect the effort of the other controllers.

The above control strategy may be extended to the control of multiple modes. Again, for ease of illustration, consider that two modes, e.g., modes #1 and #2, are controlled by four controllers, respectively. In particular, the controllers at $u_1$ and $u_2$ access the modal signal of mode #1, the controllers at $u_3$ and $u_4$ access the modal signal of mode #2. The modified modes can be obtained by solving the eigenvalues of a subset of the closed-loop system state matrix, denoted as

$$\boldsymbol{A}_s^* = \begin{bmatrix} \lambda_1 + a_{11} & a_{12} \\ a_{21} & \lambda_2 + a_{22} \end{bmatrix} \quad (10)$$

where

$a_{11} = \boldsymbol{n}_1^T\boldsymbol{B}_1\boldsymbol{O}_1 + \boldsymbol{n}_1^T\boldsymbol{B}_2\boldsymbol{O}_2;\ a_{12} = \boldsymbol{n}_1^T\boldsymbol{B}_3\boldsymbol{O}_3 + \boldsymbol{n}_1^T\boldsymbol{B}_4\boldsymbol{O}_4;$
$a_{21} = \boldsymbol{n}_2^T\boldsymbol{B}_1\boldsymbol{O}_1 + \boldsymbol{n}_2^T\boldsymbol{B}_2\boldsymbol{O}_2;\ a_{22} = \boldsymbol{n}_2^T\boldsymbol{B}_3\boldsymbol{O}_3 + \boldsymbol{n}_2^T\boldsymbol{B}_4\boldsymbol{O}_4.$

Ideally, if the off-diagonal terms of (10) are inherent zeros, the controllers for the same mode and different modes are fully decoupled, because the eigenvalues of $\boldsymbol{A}_s^*$ are given by its diagonal terms. But in reality, the off-diagonal terms may be non-trivial complex numbers. Moreover, it is noted that the $O$s in (10) are non-trivial numbers as they are designed to modify the modes. So, to apply the proposed strategy to control both modes, $\boldsymbol{A}_s^*$ should be at least nearly either a lower or upper triangular matrix; then, the eigenvalues of $\boldsymbol{A}_s^*$ can be approximately given by its diagonal terms. To this end, the controllers that target mode #2 may need to be enabled at places that have small controllability for mode #1; then the cross modal terms induced by the control of mode #1 is no longer crucial in terms of modal interactions. Moreover, if the natural frequency of mode #1 is close to the frequency of mode #2, then both $a_{12}$ and $a_{21}$ may need to be small, which requires that the control sites for mode #1 also have small controllability for mode #2.

*B. Application to Multi-Point Active Power Modulation*

Active power modulation may be executed by a system device that has fast power response capability, e.g., energy storage system. The dynamics of the power modulation may be modeled as a first order transfer function [8], as depicted in Fig. 2. When a controller in Fig. 1 is connected to $P_{ref}$, as the system is excited, the device will provide a stabilizing signal $\Delta P_{in}(t)$ to the system[1].

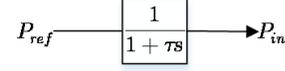

Fig. 2. Simplified active power modulation block.

When applying the proposed strategy to control a group of such devices to damp a particular mode of frequency oscillation (electromechanical or common low-frequency), the power modulations may be engineered by employing the following method. For a particular mode of oscillation, because the outputs of the controllers are in proportion to their respective $O_d$, to pursue consistent amplitude of power response $|\Delta P_{in}|$ for all the devices, the gains of the controllers may be set up according to

$$|\boldsymbol{n}_j^T\boldsymbol{B}_1\boldsymbol{O}_1|/|\boldsymbol{n}_j^T\boldsymbol{B}_2\boldsymbol{O}_2| = |\boldsymbol{n}_j^T\boldsymbol{B}_1|/|\boldsymbol{n}_j^T\boldsymbol{B}_2| \quad (11)$$

Furthermore, to impose constraints on the response of each device, choose

$$|\boldsymbol{n}_j^T\boldsymbol{B}_1\boldsymbol{O}_1|/|\boldsymbol{n}_j^T\boldsymbol{B}_2\boldsymbol{O}_2| = (\varepsilon_1|\boldsymbol{n}_j^T\boldsymbol{B}_1|)/(\varepsilon_2|\boldsymbol{n}_j^T\boldsymbol{B}_2|) \quad (12)$$

where $\varepsilon$ may be the response limit of the device, denoted by $P_{max} - P_{ref}$.

The actual power responses of the devices are determined by the closed-loop system dynamics. Larger gains of the controllers tend to result in larger power responses of the devices. However, it also means larger damping will be added to the target mode, which renders a mitigation effect on the maximum values and duration of the power responses. Moreover, the concerned mode of oscillation is usually triggered by a transient event occurs in the system. In this regard, stimulation of the system at different locations in a non-linear simulation environment may be necessary to properly set up the gains of the controllers. The main tuning involved will be adjusting the weights in (12) and simultaneously scaling the gains of the controllers. Note that the damping that the controllers can provide is mainly dependent on the availability of the supporting devices and how large the oscillation will be under the most severe contingency. As more devices partake in the control of a mode, the damping contribution needed from each device can be smaller, therefore smaller pressure put on the individual devices.

For a large power system with geographically dispersed actuators, the multi-mode control strategy may be applied. This is based on the fact that the electromechanical modes are usually controllable at some particular sites (areas) and the common low-frequency mode is widely controllable at the system interconnection buses. So it is possible to group the devices to simultaneously control multiple modes with less control interactions.

---

1. This type of modulating control is physically interpretable, it has been well observed from the experiments that a qualified control signal tends to be out of phase with the oscillation signal seen by the device's terminal bus frequency, which indicates that the device will dissipate the oscillation energy or say provide damping torque to the oscillation.

## IV. Determination of the Input Signal

In this section, the LQRSP technique is formalized as a tool for determining the input signal of the controller depicted in Fig. 1, some practical considerations of the presented approach are also discussed. Then, the procedure for controller design is summarized.

### A. Determination of the Input Signal Using LQRSP

In (9), the proposed damping controller needs to access a pure modal signal to achieve the respective mode mobility. But it may be difficult to realize this in practice. Therefore, the goal here is to help the controllers access a modal selective signal whose modal observability is dominated by the target mode. In static feedback control, this may be accomplished by carefully pairing the system outputs and inputs. For instance, using $\sum_q K_{d,q} C_q x$ to obtain an input signal, which features $\left|\sum_q K_{d,q} C_q m_j\right| \gg \left|\sum_q K_{d,q} C_q m_r\right|$ for any mode $r$, for the purpose of improving the damping of the $j^{th}$ mode. Nevertheless, it is not easy to achieve this for a large power system without the help of a systematic algorithm, as each system state may significantly participate in multiple mode dynamics; see (2). Ideally, we expect a solution so that the required system outputs are readily accessible via dynamic state estimators [14] or phasor measurement unit (PMU) values, e.g., the rotor speed. As follows, the LQRSP technique [15] is formalized as a tool for determining the desired input signal using the least number of system states.

*a) Brief Review of the LQRSP:* If we consider that the system output in (1b) is a combination of full system states, i.e., the state vector is observed by an $n \times n$ identity matrix, the LQR-based sparsity-promoting optimal control problem can be formulated as [16]:

$$\begin{cases} \text{minimize } \lim_{t\to\infty} \varepsilon\{x(t)^T Q x(t) + u(t)^T R u(t)\} \\ \quad\quad + \gamma_{i,j} \sum_{i,j} w_{i,j} |F_{i,j}| \\ \text{subject to} \\ \text{dynamics: } \dot{x}(t) = Ax(t) + B_{s1} u(t) + B_{s2} \eta(t) \\ \text{linear control: } u(t) = -Fx(t) \\ \text{stability: the real parts of } eig(A - B_{s1}F) < 0 \end{cases} \quad (13)$$

where $\gamma$ is usually a small positive number within a range to promote the sparsity of the gain matrix $F$, $\varepsilon\{\bullet\}$ is the expectation operator, $Q$ is an $n \times n$ positive semi-definite matrix designed to penalize particular system dynamics, $R$ is a $p \times p$ positive definite matrix, and $p$ is the number of system inputs selected to equip controllers, for our distributed controller design, $p = 1$. Normally, the matrix $R$ is an identity matrix, and $B_{s1} = B_{s2}$. For $\gamma = 0$, the control reduces to a standard $H_2$ optimal control problem, the optimal gain matrix $F_c$ can be obtained by solving the Algebraic Riccati Equation (ARE), and $F_c$ is usually fully populated with non-zero entries. By gradually increasing $\gamma$, the sparsity promoting process tends to strike a balance between the dynamic response performance and the number of communication links (i.e., the number of non-zero entries of $F$). The algorithm is available in Matlab as *lqrsp*.

*b) LQRSP as a Tool for Determining the Modal Selective Signal:* In [16], the authors mentioned that the sparse LQR may be modal selective by using a proper state cost function, e.g., the state cost can be squared relative rotor speed and angle between two groups of machines for the purpose of controlling a specific inter-area mode. Thus inspired, two types of state costs are employed below to search the desired input signals for the controllers.

For the control of an electromechanical mode, choose

$$x^T Q x = \zeta(\omega_\alpha - \omega_\beta)^2 \quad (14)$$

And for the control of the common low-frequency mode, choose

$$x^T Q x = \zeta(\omega_{\text{COI}})^2; \quad \omega_{\text{COI}} = \frac{\sum_{\alpha=1}^{n_m} H_\alpha \omega_\alpha}{\sum_{\alpha=1}^{n_m} H_\alpha}. \quad (15)$$

where $\omega_\alpha$ and $\omega_\beta$ denote a pair of (aggregated) synchronous machine speed variables that oscillate against each other under a particular inter-machine oscillation, $n_m$ is the amount of synchronous machine in the system, and $\zeta$ is a scalar; $\omega_{\text{COI}}$ denotes the center-of-inertia (COI) speed of the system, $H_\alpha$ and $\omega_\alpha$ denote respectively the inertia time constant and the speed variable of synchronous machine $\alpha$.

*c) Interpretation:* Given that the LQR can be viewed as a controller that intends to shrink the energy of the input-output frequency response [17], the following inference may serve as an interpretation of the approach. Based on (3a), (3b) and the modal properties, and assuming zero initial conditions, a unit impulse in the selected input yields the following responses in the $i^{th}$ output

$$y_i(t) = \sum_{j=1}^n C_i m_j n_j^T B_{s1} e^{\lambda_j t} \quad (16)$$

Therefore, a particular mode may dominate a synthesized output response by linearly combining the responses of different system outputs. Using (14), the response of the common mode tends to be excluded while the response of a particular inter-machine mode is emphasized. Using (15), the responses of inter-machine modes are greatly eliminated while the response of the common mode is retained as the speed variables oscillate coherently under the common mode dynamics. According to (6) and (7) and following the logic of the LQR control, it can be inferred that if the impulse response of a system output is dominated by a single mode, then the controller tends to find a gain matrix that stresses the modal signal of the dominant mode while excluding the other modal signals. Because excessive modal signals of non-dominant modes will increase the control energy yet are unable to reduce the response energy of the dominant mode, and hence the overall energy, the result is the desired input signal.

*d) Practical Considerations:* In practice, the modal selectivity of the input signal tends to be perfect when using the optimal gain matrix. The interaction among the controllers, which violates the decoupling goal, tends to increase as the modal selectivity deteriorates. Therefore, a trade-off occurs between the desired control performance and the sparsity of the gain matrix for each controller. As confirmed later, the presented approach helps to find the desired input signal for a controller resulting from a linear combination of the states of the systems. And, thanks to the sparsity-promoting technique, all the controllers need to access rotor speed variables of the synchronous generators but without sacrificing the optimal control performance possibly because of the modal structure of the modes. In addition, the selection of $\alpha$ and $\beta$ in (14) is the result of modal structure investigation and coherency grouping, prior knowledge of the system is preferred.

*B. Damping Controller Design Flowchart*

Overall, the design flowchart of the damping controllers for a particular mode is depicted in Fig. 3. Steps 1-2 aim to prepare the linear model and choose control locations for the concerned mode. Steps 3-4 are related to the determination of the input signal for each controller using the approach introduced in Section IV-A; as usual in control design, trial and error might be encountered especially for the parameters used in the LQRSP algorithm. Steps 5-6 are devoted to the gain tuning using the approach introduced in Section III-B.

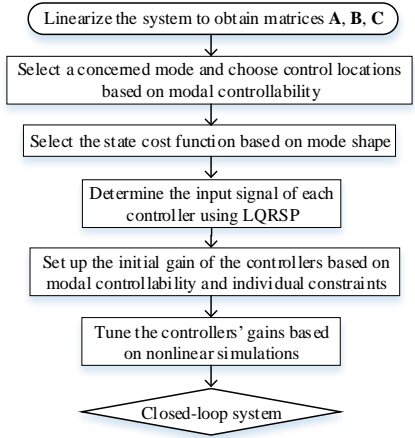

Fig. 3. Design flowchart of the damping controllers for a particular mode.

## V. Validation on a Two-Area System

*A. System Description and Problem Statement*

Kundur's two-area system in [18] is adapted for a detailed validation of the proposed methodology. Two active power injection devices are integrated into the system near the loads as depicted in Fig. 4. The devices are both operated at 20 MW. For illustration purposes, there is no response limit assigned to the devices. The generators are modeled using a sub-transient generator model (6th order). Simple exciter (1st order) is installed on the generators. Simple turbine-governor model (3rd order) is employed to model the mechanical power dynamics of the generators. No PSS is enabled. The loads are modeled as constant impedance loads for both the active and reactive power consumptions. Eigen-analysis reveals that the system is stable, and the electromechanical modes and the common mode are listed in Table I. It is seen that the system's inter-area mode is lightly damped due to possibly the heavy power transfer and the fast-acting voltage regulators. The goal here is to damp the inter-area oscillation by efficiently modulating the active power injections of the integrated devices. Before proceeding, to gain some insights for the necessity of a scientific control design, some empirical output feedback strategies are reviewed. For a static feedback control loop that is established between the specified system output and the power set point of D2, the root locus is depicted in Fig. 5. Although all of the strategies can improve the damping of the inter-area mode, several important issues are noteworthy:

1) When using the rotor speed of generator 2 as control input, the inter-area mode obtains relatively small mode mobility compared to the common mode and local mode 1. So, to achieve a desired damping improvement of the inter-area mode, excessive control energy may be consumed by those two well-damped modes.
2) The 'local average speed' $\omega_1 + \omega_2$ removes the local mode mobility, but the common mode mobility is still relatively large.
3) The 'local speed difference' $\omega_1 - \omega_2$ destabilizes the local mode while adding small damping to the inter-area mode. In addition, $\omega_2 - \omega_1$ stabilizes local mode 1 while destabilizing the inter-area mode.
4) The 'speed difference' $\omega_1 - \omega_3$ slightly degrades the damping of local mode 1. Therefore, the damping improvement of the inter-area mode needs to be limited in order to alleviate the side-effect on the local mode.

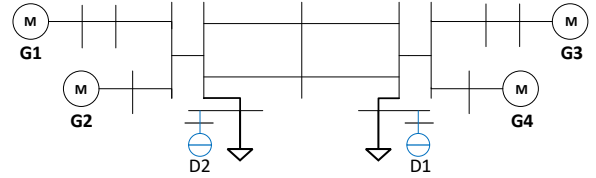

Fig. 4. Two-area system. D1 and D2 are the integrated power devices.

Table I
Electromechanical Modes and the Common Mode

| Modes | Eigenvalues |
| --- | --- |
| Local 1 (G1 vs G2) | –0.5344 + 7.2711i |
| Local 2 (G3 vs G4) | –0.8146 + 8.0112i |
| Inter-area (G1,G2 vs G3,G4) | **–0.0414 + 4.1733i** |
| Common (Coherent) | –0.4720 + 0.3909i |

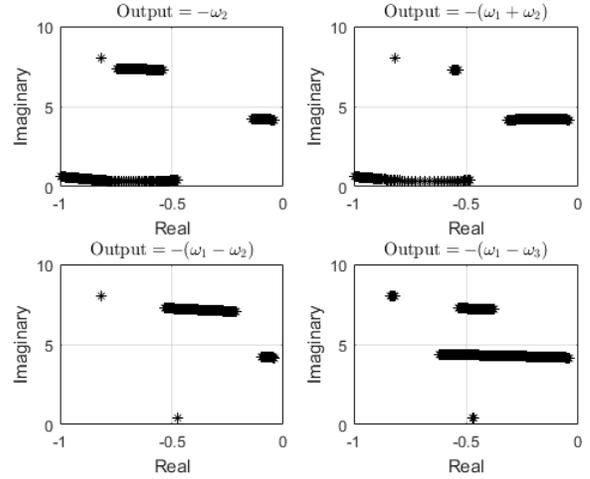

Fig. 5. Root locus, two-area system case.

The similar results are observed for D1. Moreover, the movement of system modes on the complex plane may be more complicated when multiple devices partaking in the control of a more complex system. In the following, the proposed method is first applied, and then it is compared with the optimal control strategy in [16], which suggests a feedback that controls both the inter-area mode and the common mode. The usefulness of the proposed method can be illustrated from this comparison.

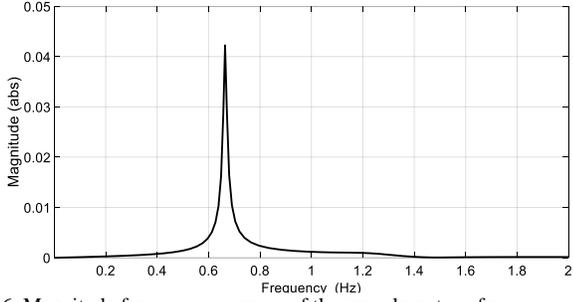
Fig. 6. Magnitude frequency response of the open-loop transfer.

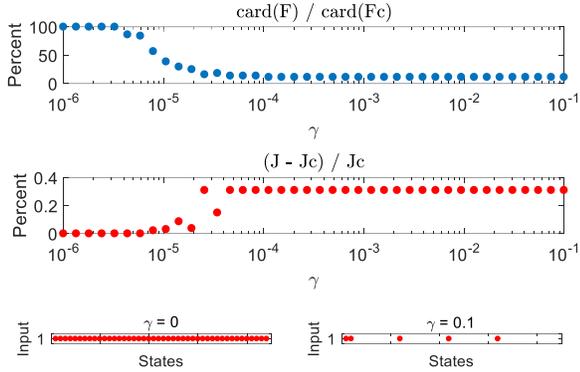
Fig. 7. Sparsity-promoting results. card = cardinality. $(J - J_c)/J_c$ denotes the quadratic performance degradation of a sparse gain matrix $\boldsymbol{F}_s$ relative to the optimal gain matrix $\boldsymbol{F}_c$.

## B. Proposed Solution

For illustration purposes, the design process of controller 1 (C1) for D1 is detailed here. $\zeta(\omega_1 - \omega_3)^2$ is chosen as the state cost for the LQR, where $\omega_1$ and $\omega_3$ denote respectively the rotor speed variables of generators 1 and 3. The magnitude frequency response of the open-loop transfer between the device's power set point and the synthesized output $\omega_1 - \omega_3$ is shown in Fig. 6. The inter-area mode dominates the response, so it is expected that the resulting gain matrix of the LQR stresses the modal signal of the inter-area mode. By using $\gamma = [10^{-6}, 10^{-1}]$ with 40 logarithmically spaced values, the LQRSP delivers following information. As shown in Fig. 7, for $\gamma = 0$, the gain matrix given by ARE is fully populated and, as $\gamma$ increases, the number of non-zero entries of the gain matrix is greatly reduced while the control performance slightly deteriorates. For $\gamma = 0.1$, the gain vector is $\boldsymbol{F}_{\gamma=0.1} = [0.0002, -1.1139, \boldsymbol{0}, -1.0413, \boldsymbol{0}, 1.1873, \boldsymbol{0}, 1.0431, \boldsymbol{0}]$, where $\boldsymbol{0}$ denotes zeros, the four relatively large numbers are the gains for the rotor speed variables of generators 1-4, respectively. The relatively small number is the gain for the rotor angle state of generator 1. It is very small, thus it is reasonable to remove it from the gain vector. Therefore, the gain vector selected for C1 is $\boldsymbol{F}_s = [0, -1.1139, \boldsymbol{0}, -1.0413, \boldsymbol{0}, 1.1873, \boldsymbol{0}, 1.0431, \boldsymbol{0}]$, which means that the input signal of the controller in Fig. 1 is $1.1139\,\omega_1(t) + 1.0413\,\omega_2(t) - 1.1873\,\omega_3(t) - 1.0431\,\omega_4(t)$. The ideal closed-loop system modes given by $eig(\boldsymbol{A} - 58\boldsymbol{B}_{D1}\boldsymbol{F}_s)$ are depicted in Fig. 8; #58 is the controller gain. It can be seen that only the inter-area mode is significantly modified indicating the modal selectivity of the input signal. To verify the theoretical derivation, the mode-move calculated by the scaled mode mobility in (9) is given in (17).

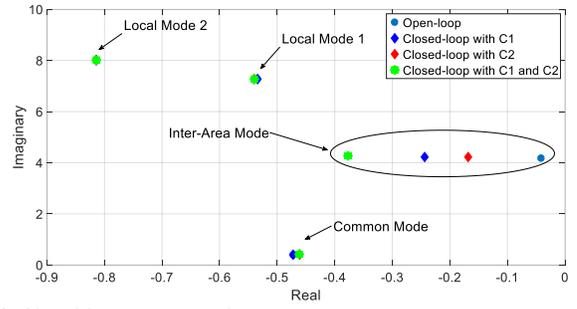
Fig. 8. Closed-loop system modes.

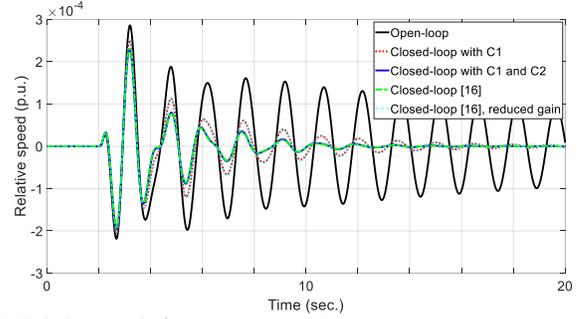
Fig. 9. Relative speed of generators 1 and 3.

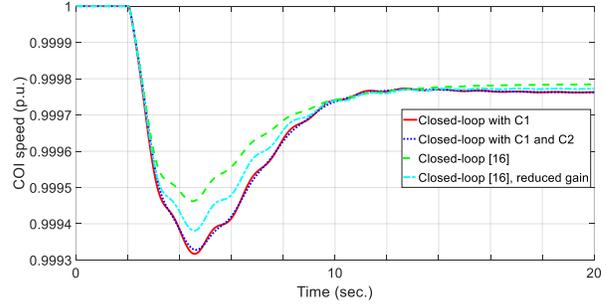
Fig. 10. COI speed.

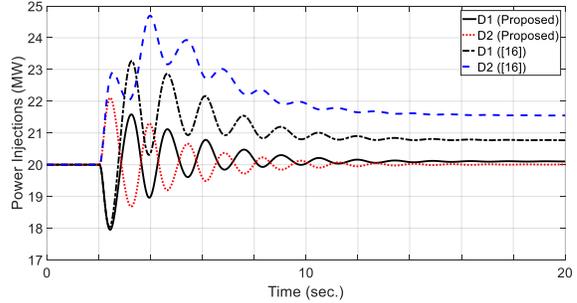
Fig. 11. Power responses of the devices. Proposed method vs [16].

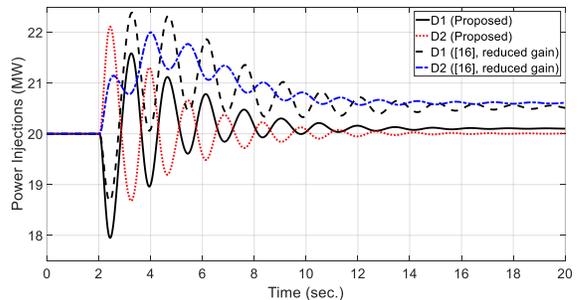
Fig. 12. Power responses of the devices. Proposed method vs [16] with reduced gain.

$$n_{inter}^T B_{D1} \times 58(1.1139 C_{spd1} m_{inter} + \\ 1.0413 C_{spd2} m_{inter} - 1.1873 C_{spd3} m_{inter} - \\ 1.0431 C_{spd4} m_{inter}) = -0.1973 + j0.0531 \quad (17)$$

The controller 2 (C2) for D2 is designed by following the same procedure, and the gain of the controller is tuned based on (11) using C1 as a reference. So the devices are expected to have a similar amplitude of power response during the modulation process. The modes of the closed-loop system with C2 and that with two controllers are also depicted in Fig. 8, respectively.

Non-linear simulations are carried out to show the impact of the controllers on the system's oscillatory dynamics. At 2 s, the system is excited by a step increase in the exciter reference of generator 2. Under such type of disturbance, the system will undergo a process of converging to a new synchronous reference. This however alters the linear time invariant system assumption that the control approach relies on. Nevertheless, such disturbance provides an opportunity to examine the performance of the controllers when the synchronous reference varies within a small range, which is the nature of a real power system. In this paper, the rotor speed variables are captured by differentiating each rotor speed from a common reference (i.e., 1 p.u.). The relative speed of generators 1 and 3 is plotted in Fig. 8. It evidently shows that the controllers work in a superposition manner to damp the inter-area oscillation. Fig. 9 shows the devices' power responses during the modulation process, which indicates that only the oscillation signal of the inter-area mode is passing thru the controllers and the gain tuning for shaping the power responses is effective. The COI speed in Fig. 10 shows that the controllers have almost no influence on the trajectory of the system's synchronous reference. Indeed, as shown in Fig. 8, the controllers have trivial influence on the system's common mode, which defines the shape of a small-signal frequency swing [19].

### C. Comparison with the Optimal Control Strategy in [16]

The power devices are enabled for implementing a sparse LQR with the optimal control strategy in [16], i.e., the $R$ in (13) is now a $2 \times 2$ identity matrix; and the state cost is $x^T Q x = \zeta(\sum_{\alpha=1}^{4} \omega_\alpha^2)$. Sparsity-promoting results show that the controller tends to access the four generator speed variables without sacrificing the optimal control performance. The resulting controller is tested by the same disturbance. As shown in Figs. 9-11, the controller takes more control energy from the devices to achieve the same damping improvement for the inter-area mode as the proposed controllers. This is because some control energy are spent on controlling the common mode; however, the improvement of the system's primary frequency response is not very significant besides the frequency nadir. Moreover, if the devices are not ready to permanently increase/decrease their power injections (e.g., due to economic dispatch), the control actions will be infeasible. By reducing accordingly the gain for the power command signal for each device, as shown in Fig. 12, the maximum power responses are aligned with that commanded by the proposed controllers; however, there are still non-trivial steady-state bias, and the effect that the controller has on the inter-area mode is reduced, as shown by the relative generator speed in Fig. 9 and the tails of the power responses in Fig. 12.

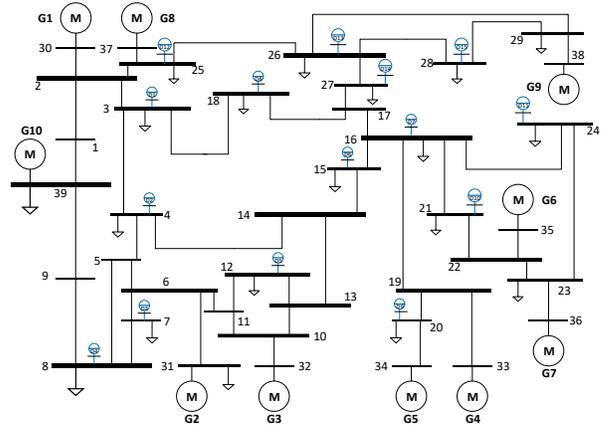

Fig. 13. New England 39-bus system. Blue circles denote the integrated active power injection devices.

### D. Discussions

Based on the above study, the advantages of the proposed control strategy may be briefly summarized as follows: It shows improved flexibility and efficiency than the conventional strategy in terms of mitigating the critical oscillation with bounded stabilizing signals. Because of the (verified) superposition feature, the single-point control energy can be greatly reduced if considerable devices partake in the control, thereby reducing the effects of control actions on the steady-state operation of the devices. These features may be favorable for the utilization of DERs in wide-area damping control.

## VI. SIMULATION ON A LARGE SYSTEM

To examine the applicability and scalability of the control approach, a study on the New England 39-bus system is carried out. 15 active power injection devices are integrated into the system, as depicted in Fig. 13. For illustration purposes, there is no response limit assigned to the devices. The power export of the NE system, i.e., $P_{1-39} + P_{9-39}$, exhibits a large swing under a power imbalance event; careful examination reveals that the swing is mainly governed by the 0.6 Hz inter-area mode and the system's common mode. The goal here is to mitigate the swing of this key power transfer by commanding the power injections of the 15 devices with the proposed control strategy (a large swing may trigger the protection and therefore the cascading issues). According to the modal controllability shown in Fig. 14, devices {D1, D5–D15} are selected to improve the damping of the 0.6 Hz inter-area mode, the state cost is $x^T Q x = \zeta(\frac{1}{9}\sum_{\alpha=1}^{9} \omega_\alpha - \omega_{10})^2$; devices {D2–D4} are selected to improve the damping of the common mode and equation (15) is employed as the state cost. After determining the input signals (only rotor speed variables are involved), the gains of the controllers are initially tuned based on (11) and then simultaneously scaled to achieve a desired damping for the respective mode. The root locus is depicted in Fig. 15. As shown in Figs. 16 and 17, in response to a sudden load increase in the NE part of the system, the devices work independently to damp their respective mode of oscillations. As a result, the swing of the power export of the NE system is reduced by about (peak-to-peak) 50 MW for this particular event. The power responses of {D2-D4} are large; however, the pressure of each device can be relieved if more devices nearby are responsive.

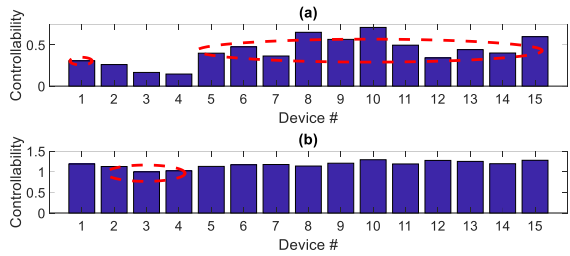

Fig. 14. Modal controllability of the devices. (a) 0.6 Hz inter-area mode. (b) The common mode. Selected devices for controlling each mode are encircled.

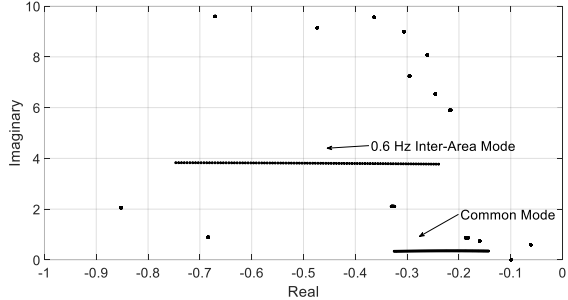

Fig. 15. Root locus, New England 39-bus system case.

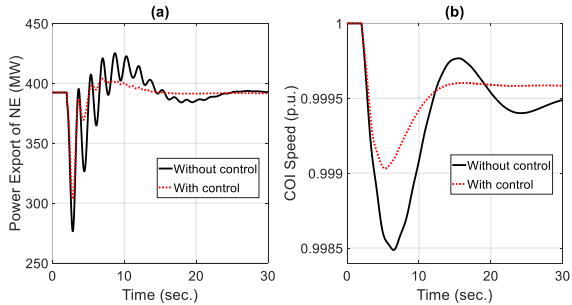

Fig. 16. (a) Power export of the NE system. (b) COI speed.

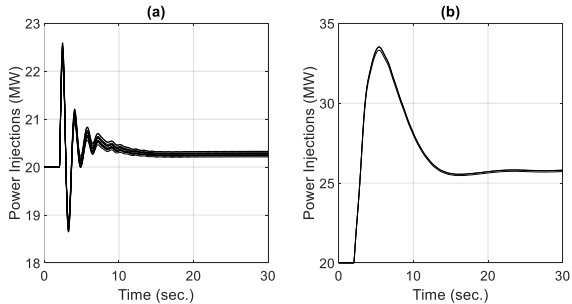

Fig. 17. Power responses of the devices. (a) {D1, D5–D15}. (b) {D2–D4}.

## VII. CONCLUSIONS AND FUTURE WORK

This paper proposes a new wide-area control strategy for modulating multiple active power injections to mitigate the critical frequency oscillations. The initial results showed that the approach has the potential to be a choice for system planners to unlocking large-scale DERs for damping control under particular operating conditions, with the unique advantage to distribute the control effort to many sites so that small modulation is needed at each single site.

However, at the current stage, the implementation of the strategy still relies on large volumes of data scattered in the system. An interesting topic of the future work is to develop a method to extract the modal oscillation signal from the local frequency measurement as the input signal of a controller may come from a single system output; see (8). Moreover, the feasibility of mitigating a sustained frequency oscillation by modulating the fast active power injections using the filtered local frequency deviations as commands may need to be examined.

## VIII. REFERENCES

[1] C. Y. Chung, L. Wang, F. Howell, and P. Kundur, "Generation rescheduling methods to improve power transfer capability constrained by small-signal stability," *IEEE Trans. Power Syst.*, vol. 19, no. 1, pp. 524–530, Feb. 2004.

[2] I. Kamwa, R. Grondin, and Y. Hebert, "Wide-area measurement-based stabilizing control of large power systems—A decentralized/hierarchical approach," *IEEE Trans. Power Syst.*, vol. 16, no. 1, pp. 136–153, Feb. 2001.

[3] B. Chaudhuri and B. C. Pal, "Robust damping of multiple swing modes employing global stabilizing signals with a TCSC," *IEEE Trans. Power Syst.*, vol. 19, no. 1, pp. 499–506, Feb. 2004.

[4] L. Chen, X. M. Lu, and Y. Min, *et al.*, "Optimization of governor parameters to prevent frequency oscillations in power systems," *IEEE Trans. Power Syst.*," vol. 33, no. 4, pp. 4466–4474, Jul. 2018.

[5] L. Chen, Y. Min, and X. M. Lu, *et al.*, "Online emergency control to suppress frequency oscillations based on damping evaluation using dissipation energy flow," *International Journal of Electrical Power and Energy Systems*, vol. 103, pp. 414–420, Dec. 2018.

[6] B. J. Pierre, F. Wilches-Bernal, D. A. Schoenwald, R. T. Elliott, D. J. Trudnowski, R. H. Byrne, and J. C. Neely, "Design of the Pacific DC Intertie wide area damping controller," *IEEE Trans. Power Syst.*, vol. 34, no. 5, pp. 3594–3604, Sep. 2019.

[7] B. K. Poolla, D. Gross, and F. Dörfler, "Placement and implementation of grid-forming and grid-following virtual inertia and fast frequency response," *IEEE Trans. Power Syst.*, vol. 34, no. 4, pp. 3035–3046, Jul. 2019.

[8] Y. L. Zhu, C. X. Liu, K. Sun, D. Shi, and Z. W. Wang, ''Optimization of battery energy storage to improve power system oscillation damping,'' *IEEE Trans. Sustain. Energy*, vol. 10, no. 3, pp. 1015–1024, Jul. 2019.

[9] A. Delavari and I. Kamwa, "Improved optimal decentralized load modulation for power system primary frequency regulation," *IEEE Trans. Power Syst.*, vol. 33, no. 1, pp. 1013–1025, Jan. 2018.

[10] F. Wilches-Bernal, D. A. Copp, G. Bacelli, and R. H. Byrne, "Structuring the optimal output feedback control gain: a soft constraint approach," 2018 IEEE Conference on Decision and Control (CDC).

[11] F. Wilches-Bernal, R. H. Byrne, and J. M. Lian, "Damping of inter-area oscillations via modulation of aggregated loads," *IEEE Trans. Power Syst.*, to be published.

[12] S. S. Guggilam, C. H. Zhao, E. Dall'Anese, Y. C. Chen, and S. V. Dhople, "Optimizing DER participation in inertial and primary-frequency response," *IEEE Trans. Power Syst.*, vol. 33, no. 5, pp. 5194–5205, Sep. 2018.

[13] J. B. Zhang, C. Y. Chung, and Y. D. Han, "A novel modal decomposition control and its application to PSS design for damping inter-area oscillations in power systems," *IEEE Trans. Power Syst.*, vol. 27, no. 4, pp. 2015–2025, Nov. 2012.

[14] E. Ghahremani and I. Kamwa, "Local and wide-area PMU-based decentralized dynamic state estimation in multi-machine power systems," *IEEE Trans. Power Syst.*, vol. 31, no. 1, pp. 547–562, Jan. 2016.

[15] F. Lin, M. Fardad, and M. R. Jovanović, "Design of optimal sparse feedback gains via the alternating direction method of multipliers," *IEEE Trans. Automat. Control*, vol. 58, no. 9, pp. 2426–2431, Sep. 2013.

[16] F. Dörfler, M. R. Jovanović, M. Chertkov, and F. Bullo, "Sparsity-promoting optimal wide-area control of power networks," *IEEE Trans. Power Syst.*, vol. 29, no. 5, pp. 2281–2291, Sep. 2014.

[17] X. F. Wu, F. Dörfler, and M. R. Jovanović, "Input-output analysis and decentralized optimal control of inter-area oscillations in power systems," *IEEE Trans. Power Syst.*, vol. 31, no. 3, pp. 2434–2444, May 2016.

[18] P. Kundur, Power System Stability and Control. New York, NY, USA: McGraw-Hill, 1994.

[19] R. C. Xie, I. Kamwa, D. Rimorov, and A. Moeini, "Fundamental study of common mode small-signal frequency oscillations in power systems," *International Journal of Electrical Power and Energy Systems*, vol. 106, pp. 201–209, Mar. 2019.